\documentclass[onecolumn,showpacs,nobibnotes,nofootinbib,12pt]{revtex4-1}
\usepackage{amsmath,amssymb}
\usepackage{wasysym}
\usepackage{graphicx}
\newcommand{\be}{\begin{equation}}
\newcommand{\ee}{\end{equation}}
\newcommand{\bea}{\begin{eqnarray}}
\newcommand{\eea}{\end{eqnarray}}
\newcommand{\f}{\frac}

\newcommand{\nn}{\nonumber}

\newcommand{\RE}{\mathbf{Re}~}
\begin{document}
\author{Bin Wu}
\email{Bin.WU@cea.fr}
\affiliation{Institut de Physique Th\'{e}orique, CEA Saclay, 91191, Gif-sur-Yvette Cedex, France}

\title{On holographic thermalization and gravitational collapse of tachyonic scalar fields}
\begin{abstract}
In this paper we study the thermalization of a spatially homogeneous system in a strongly coupled CFT. The
non-equilibrium initial state is created by switching on a relevant perturbation in the CFT vacuum during $\Delta t\gtrsim t\gtrsim -\Delta t$. Via AdS/CFT, the thermalization process corresponds to the gravitational collapse of a tachyonic scalar field ($m^2 = -3$) in the Poincare patch of $AdS_5$. In the limit  $\Delta t < \f{0.02}{T}$, the thermalization time $t_T$ is found to be quantitatively the same as that of a non-equilibrium state created by a marginal perturbation discussed in Ref. \cite{PaperI}. In the case $\Delta t \gtrsim \f{1}{T}$, we also obtain double-collapse solutions but with a non-equilibrium intermediate state at $t = 0$. In all the cases our results show that the system thermalizes in a typical time $t_T \simeq \f{O(1)}{T}$. Besides, a conserved energy-moment current in the bulk is found, which helps understand the qualitative difference of the collapse process in the Poincare patch from that in global $AdS$\cite{Bizon01,Bizon02}.
\end{abstract}
\maketitle
%
%
%
\section{Introduction}
In the AdS/CFT proposed by Maldacena\cite{Maldacena}, a complete understanding of the $\mathcal{N} = 4$ SYM (CFT) requires the information of the whole infinite tower of massive Kaluza-Klein modes on $AdS_5\times S_5$ as well as the low energy supergravity on $AdS_5$\cite{Witten}. Massless, massive and tachyonic fields in $AdS_5$ respectively correspond to marginal, irrelevant and relevant perturbations in the boundary CFT. Thermalization of the strongly coupled system in CFT may be studied holographically by reconstructing the bulk geometry from the back-reaction of those matter fields\cite{holographicReconstruction}.

The thermalization of a spatially homogeneous system initially in a non-equilibrium state created by a marginal perturbation has been studied both perturbatively\cite{Minwalla} and numerically\cite{PaperI}. The perturbation is excited by a scalar source coupled to a marginal composite operator. On the AdS side, one needs to study the black hole formation from the gravitational collapse of a massless scalar field in the bulk.  It was shown that such a strongly coupled system thermalizes in a typical time $t_T\sim\f{O(1)}{T}$ with $T$ being the thermal-equilibrium temperature\cite{PaperI}. $t_T$ is the typical time for the scalar field to fall from the boundary to the interior of $AdS_5$ close to the location of the apparent horizon. For the other aspects of the thermalization of such a spatially homogeneous system, the interested reader is referred to \cite{Esko:1999,LinShuryak, holographicThermal,Bizon01,Bizon02,Garfinkle01,Garfinkle02,Galante:2012pv,Baier:2012a,Baier:2012b,Erdmenger:2012xu,Steineder:2012si,
Arefeva:2012jp,Muller:2012rh,Baron:2012fv,Caceres:2012px} for CFT and \cite{Gelis:2010,Gelis:2011} for classical field theory.

If the boundary source is switched on only in a short duration, one may expect that the thermalization time $t_T$ should not (dramatically) depend on whether it is coupled to a marginal or relevant operator. However, on the gravity side the thermalization processes in these two cases respectively correspond to the gravitational collapse of massless or tachyonic scalar fields. It is interesting to see under what conditions the mass of the scalar field does not affect significantly the thermalization time.

In this paper we study the thermalization process of a spatially homogeneous system. The system is initially in a non-equilibrium state created by turning on a scalar source coupled to a relevant operator $O_{\Delta}$ in CFT. The energy is injected into the system according to\cite{holographicReconstruction}
\be
\dot{\varepsilon}\equiv{{\dot{T}}^{ (4) } }_{00} = -\left< O_{\Delta}\right> \dot{\phi_{(0)}},\label{equ:energyConservation}
\ee 
where $ \left< O_{\Delta}\right>$ is the expectation value of $O_{\Delta}$ and  the overdots denote time derivatives. After the source is switched off, the system will finally thermalize through self-interaction and the information about the initial state will be totally lost. In $AdS_{d+1}$, it corresponds to the gravitational collapse of a tachyonic scalar field of mass $m$. Stability requires not $m^2 >0$ but $m^2 \geq -\f{d^2}{4}$. The conformal dimension $\Delta$ of $O_{\Delta}$ is related to the mass of the scalar field by\cite{Witten}
\be
\Delta (\Delta-d) = m^2,~~\text{i.e.,}~~\Delta = \f{d}{2}+\sqrt{\f{d^2}{4}+m^2} = \f{d}{2} + n
\ee
with \be n=\sqrt{\f{d^2}{4}+m^2}.\ee
The scalar source induces a tachyonic scalar wave falling from the boundary into the interior of $AdS_{d+1}$. It satisfies the boundary condition
\be
\phi(t,u)\sim u^{d-\Delta} \phi_{(0)}(t).\label{equ:phisim}
\ee
The induced wave will eventually collapse to form a black hole, which is the gravity dual of a static plasma. At the end, it will be completely hidden behind the apparent horizon. The thermalization process in CFT will be studied by investigating such a collapse process in $AdS_{d+1}$. In this paper, $d=4$ and the scalar source $\phi_{(0)}$ is chosen to take the form  
\be
\phi_{(0)}(t) = \f{\epsilon}{a} e^{-a t^2},\label{equ:phi0t}
\ee
where $\epsilon$ and $a\equiv \f{1}{\Delta t^2}$ are two parameters and $\Delta t$ characterizes the duration of the source being turned on.

This paper is organized as follows. In Sec. \ref{sec:eom} we give the equations of motion for a scalar field of arbitrary mass coupled to gravity in the Poincare patch of $AdS_5$. The numerical scheme in addition to initial conditions and boundary conditions is also discussed in this section. Then, we study the propagation of a scalar field of arbitrary mass in $AdS_5$ with the back-reaction to the bulk geometry being ignored in Sec. \ref{vacuum}. In Sec. \ref{sec:results}, we investigate the details about the gravitational collapse of the tachyonic scalar field with $m^2 = -3$. In Sec. \ref{sec:discussion}, we briefly conclude. The near-boundary behavior of the massless scalar field is presented in Appendix \ref{sec:appA}. In Appendix \ref{sec:appB}, a conserved energy-momentum current in global $AdS_4$ is defined, which is useful for understanding some details about the collapse process of scalar fields discussed in \cite{Bizon01,Bizon02}.
%
%
%

\section{ Einstein-Klein-Gordon equations}
\label{sec:eom}

\subsection{Equations of motion}
On the AdS side, we need to solve the Einstein-Klein-Gordon equations
\bea
&&\f{1}{\sqrt{-g}}\partial_a\left(\sqrt{-g} g^{ab} \partial_b \phi \right) - m^2 \phi=0,\label{equ:KleinGolden}\\
&&R_{ab}-\f{1}{2} g_{ab} R - \f{d(d-1)}{2L^2} g_{ab} = T_{ab},\label{equ:Einstein}
\eea
where $T_{ab}$ takes the form
\be
T_{ab}= 2 \partial_a \phi \partial_b \phi - g_{ab} \left[  \left(\partial \phi \right)^2 + m^2 \phi^2 \right].
\ee
In the following, we take $L=1$ and $d=4$. For the spatially homogeneous system on the boundary $M^4$, we use the Schwarzschild coordinates of the form
\be\label{equ:FeffermanGraham}
ds^2 = \f{1}{u^2}\left( -f e^{-2 \delta} dt^2 + f^{-1} du^2 + d\vec{x}^2 \right),
\ee
where $f$ and $\delta$ are functions of $t$ and $u$ only. In this coordinate system, one obtains from eqs. (\ref{equ:KleinGolden}) and (\ref{equ:Einstein}) the following equations of motion
\begin{subequations}\label{equ:eom}
\bea
&&\dot{V} = u^3 \left(\f{f e^{-\delta} P}{u^3}\right)^\prime - \f{m^2}{u^2} e^{-\delta} \phi,\label{equ:Vdot}\\
&&\dot{P} = \left( f e^{-\delta} V \right)^\prime,\label{equ:Pdot}\\
&&\dot{f} = \f{4}{3} u f^2 e^{-\delta} V P,\label{equ:fdot}\\
&&\delta^\prime=\f{2}{3} u \left(  V^2 + P^2 \right),\label{equ:deltap}\\
&&f' = \f{2}{3} u\left[ f \left(V^2+ P^2\right) + \f{m^2}{u^2} \phi^2 \right]  + \f{4}{u}\left( f-1\right) ,\label{equ:fp}
\eea
\end{subequations}
where the derivatives with respect to $t$ and $u$ are denoted respectively by overdots and primes, $P\equiv \phi^\prime$ and $V\equiv f^{-1} e^\delta \dot{\phi}$.

\subsection{Initial and boundary conditions}
Before the scalar source is turned on, the scalar field is assumed to vanish in the bulk. In this case,  at the initial time $t = t_i$ the metric functions and the scalar field are as follows
 \bea
f=1,~~\delta=0,~~V=0~~\mbox{and}~~P=0.\label{equ:initial}
\eea
The boundary condition for solving (\ref{equ:deltap}), (\ref{equ:Vdot}) and (\ref{equ:Pdot}) are respectively given by
\footnote{In our numerical calculations, we have also tried providing the boundary conditions by including higher-order terms in the power series solutions of eq. (\ref{equ:eom}) in eqs. (\ref{equ:deltaseries}), (\ref{equ:fseries}) and (\ref{equ:phiseries}) but we only find negligible difference.}
\bea
&&\delta(t,0)=0,\label{equ:deltabc}\\
&&V(t,u_{min})=-2 t \epsilon e^{-a t^2} u_{min}^{2 - n} \text{ and }P(t,\infty) = 0\text{~as~}u_{min}\to 0.\label{equ:phi0}
\eea

\subsection{Scaling symmetry}
The scaling symmetry of the equations of motion in eq. (\ref{equ:eom}) allows us to obtain solutions from a known solution as follows\cite{PaperI}
\be
\phi_{\lambda}(t,u) = \phi(\lambda t,\lambda u),~f_{\lambda}(t,u) = f(\lambda t,\lambda u),\text{~and~}\delta_{\lambda}(t,u) = \delta(\lambda t,\lambda u),\label{equ:scalingTransform}
\ee
where $(\phi, f, \delta)$ denote the solution with the boundary condition given by
\be
\phi(t,u_{min})=\f{\epsilon}{a} e^{-a t^2} \left(u_{min}\right)^{2 - n}
\ee
and  $(\phi_{\lambda}, f_{\lambda}, \delta_{\lambda})$ with $\lambda>0$ denote the one with
\be
\phi(t,u_{min})=\f{\epsilon}{a} e^{-a (\lambda t)^2} \left(\lambda u_{min}\right)^{2 - n}=\f{\lambda^{4-n}\epsilon}{\lambda^2 a} e^{-(\lambda^2 a) t^2} u_{min}^{2 - n}.
\ee
Or, equivalently,
\be(a, \epsilon)  \to \left(\lambda^2 a, \lambda^{4 - n} \epsilon \right).\label{equ:parametersScaling} \ee
As a result, we only need to study the dependence of solutions either on $a$ or $\epsilon$. 

\subsection{Numerical scheme}
The equations of motion in eq. (\ref{equ:eom}) can be solved by the numerical scheme described in detail in Ref. \cite{PaperI}: Given $f$, $\delta$, $V$ and $P$ at $t = t_j$, we first calculate $V$, $P$ and $f$ at the next time step $t_{j+1}$ by solving (\ref{equ:Vdot}), (\ref{equ:Pdot}) and (\ref{equ:fdot}). Then, $\delta$  at $t_{j+1}$ is obtained by solving eq. (\ref{equ:deltap}). Given the initial conditions in (\ref{equ:initial}), the bulk metric at late times can be calculated by repeating the above two steps. In our numerical calculation, we use a grid in the $u$ direction with uniform spacing $du$. We find that a stable algorithm requires the time step $dt\lesssim 0.1 du$. Also, $du$ should satisfy $du \lesssim 0.1 u_{min}$ to ensure that the boundary conditions can be correctly provided.

For arbitrary mass, we can not in general provide the boundary condition at $u = 0$ due to our choice of the coordinates in eq. (\ref{equ:FeffermanGraham}) (see eq. (\ref{equ:phisim})). 
In this paper, the boundary condition for numerical simulations is provided at $u=u_{min}\ll1$ in eq. (\ref{equ:phi0}). We find that our numerical results are insensitive to the choice of $u_{min}$ as long as $u_{min}\leq 0.01$. For all the numerical results in the following sections, we choose $u_{min}=0.001$. 

%
%
%
\section{Scalar fields in $AdS_5$}\label{vacuum}
In this section we ignore the back-reaction of the scalars to the bulk geometry. In this case, the $AdS_5$ metric, with $f = 1$ and $\delta = 0$ in eq. (\ref{equ:FeffermanGraham}), has a killing vector
\be
\xi^a = \delta^a_0,
\ee 
which allows us to define a conserved total energy $M(t, u)$ as follows
\footnote{That is, we have \be0 = \int dx^{5} \sqrt{-g} \bigtriangledown_a \left( T^{ab}\xi_b \right) =  \int d^3x\left[\left.\int_u^\infty \f{du}{u^4} t_a \xi_b T^{ab}\right|^t_{-\infty} + \left.\int_{-\infty}^t \f{dt}{u^4} u_a \xi_b T^{ab}\right|^{\infty}_{u}  \right]\ee with $u_a \equiv \f{\delta_{au}}{u}$ and $t_a \equiv \f{\delta_{a0}}{u}$. }
\be
\f{\partial}{\partial t} M(t, u) \equiv \f{\partial}{\partial t} \int_u^\infty du \mathcal{E}^0 = \left. \mathcal{E}^u\right|^{u}_{\infty},\label{equ:energyCon}
\ee
where
\bea
 \mathcal{E}^0 \equiv \f{2}{3u^3} \left( V^2 + P^2 +\f{m^2}{u^2} \phi^2 \right),~\text{and}, ~\mathcal{E}^u \equiv -\f{4}{3u^3} PV =  -\f{4}{3u^3} \phi'\dot{\phi}.\label{equ:E0Euvac}
\eea 
In Sec. {\ref{sec:results}}, we shall show that  in the case $\f{\epsilon}{a} \ll 1$ $M(t, 0)$ is a very good approximation to the black hole mass resulted from the gravitational collapse of the scalar fields.

In such a fixed gravitational background, the equation of motion of the scalar field in eq. (\ref{equ:KleinGolden}) or eq. (\ref{equ:Vdot}) reduces to
\be
\ddot{\phi} -\phi'' + \f{m^2}{u^2}\phi +\f{3}{u}\phi'=0. 
\ee
The general solution to the above equation is given by
\be
\phi = \RE u^2 \int\f{d\omega}{2\pi}e^{-i \omega t} \left[ C_J(\omega) J_n\left(\omega u \right)+  C_Y(\omega) Y_n\left(\omega u \right)  \right],
\ee
where $J_n$ and $Y_n$ are respectively the Bessel functions of the first and second kind. For the boundary condition in eq. (\ref{equ:phisim})
\be
\phi = \RE u^2 \int\f{d\omega}{2\pi}e^{-i \omega t} C_Y(\omega) Y_n\left(\omega u \right)\label{equ:phivac}
\ee
with \be C_Y(\omega) = -\f{\pi \omega^n}{2^n \Gamma(n)}\phi_{(0)}(\omega)\text{~and~}\phi_{(0)}(\omega) \equiv \int dt e^{i\omega t} \phi_{(0)}(t) = \f{\epsilon \pi^{\f{1}{2}}}{a^{\f{3}{2}}} e^{-\f{\omega^2}{4a}}.\ee
Taking the explicit form of $\phi_{(0)}$ in eq. (\ref{equ:phi0t}), we have
\be
C_Y(\omega) = -\f{\epsilon  \omega^{n} e^{-\f{\omega^2}{4a}}}{2^{n}\Gamma(n)}\left( \f{\pi}{a}\right)^{\f{3}{2}}.
\ee

At time $t\gg \Delta t \equiv \f{1}{\sqrt{a}}$, regardless of its mass the group velocity of the scalar wave in eq. (\ref{equ:phivac}) is always equal to the speed of light. This can be proven as follows: $\phi$ is not significantly suppressed only in the region $|t-u|\lesssim \Delta t$ at late times. Using the asymptotic expansion of $Y_n(\zeta)$ at $|\zeta| \gg1$\footnote{Here, we only discuss the cases with $n$ being integral and $m^2 >-4$.}
\begin{eqnarray}
Y_n(\zeta)\simeq\left\{
\begin{array}{cc}
\f{\left(-1\right)^{l}}{\left( \pi \zeta \right)^{\f{1}{2}}} \left[ \sin \zeta -\left(-1\right)^n \cos \zeta \right],& \text{~~if~~} \zeta>0\\
\f{\left(-1\right)^{l}}{ \left( \pi \zeta \right)^{\f{1}{2}} } \left[ \left( 2 - \left(-1\right)^n i \right) \sin \zeta -   \left( i + \left(-1\right)^n 2 \right)  \cos \zeta \right],& \text{~~if~~} \zeta<0
\end{array}\right.
\end{eqnarray}
with $n = 2 l + (-1)^{n\mod2}$ in eq. (\ref{equ:phivac}),  after some algebra we obtain
\bea
\phi &\simeq& u^{\f{3}{2}}\phi_{a}\left(\f{t-u}{\Delta t}\right)\nn\\
&\equiv& \left( -1\right)^{l} C_{n} u^{\f{3}{2}} \int_0^\infty d\omega \omega^{n-\f{1}{2}}e^{-\f{\omega^2}{4}} \left[ (-1)^{n} \cos\f{\omega(t-u)}{\Delta t}+ \sin\f{\omega(t-u)}{\Delta t}\right],\label{equ:phiasy}
\eea
where
\bea
&&\phi_a(\zeta)=\left( -1\right)^{l}  C_{n} 2^{n-\f{1}{2}} \nn\\
&&\times\left[ 2 \zeta \Gamma\left(\f{3}{4} + \f{n}{2}\right) {}_{1}F_{1} \left( \f{3}{4} + \f{n}{2}, \f{3}{2}, -\zeta^2 \right) + (-1)^{n} \Gamma\left(\f{1}{4} + \f{n}{2}\right) {}_{1}F_{1} \left( \f{1}{4} + \f{n}{2}, \f{1}{2}, -\zeta^2 \right) \right]
\eea with \be
C_{n} \equiv  \f{\epsilon a^{ \f{n}{2} -\f{5}{4} } }{2^{n} \Gamma(n)}.
\ee
Here, ${}_{1}F_{1}$ is the Kummer confluent hypergeometric function. Therefore, at late times all the scalars propagate at the speed of light in the bulk.

Near the boundary $u=0$, the power series expansion of the scalar field has the form
\bea
\phi 
&=& u^{d-\Delta} \left(\phi_{(0)} + \phi_{(2)} u ^2 + \cdots + \phi_{(2n-2)} u^{2n-2} +  \phi_{(2n)} u^{2n} + \psi_{(2n)} u^{2n} \log u^2  + \cdots \right),
\eea
where $\phi_{(2m)}$ for $2m<2n$ and $\psi_{(2n)}$ are determined by the boundary source $\phi_{(0)}$. The expectation value of the dual operator $O_{\Delta}$ is given by\cite{holographicReconstruction}\be\left< O_{\Delta}\right> = 2 n \phi_{2n}\label{equ:O}\ee
up to terms contributing contact terms in the 2-point function. In the next section, we shall focus on the tachyonic scalar with $m^2 = -3$. 
In this case, $\Delta = 3$ and $n = 1$. The power series expansion of $\phi$ in eq. (\ref{equ:phivac}) near the boundary takes the form
\bea
\phi &=&  \RE u \int\f{d\omega}{2\pi} e^{-i\omega t} \phi_{(0)}(\omega)\left[1 - \f{\omega^2}{4}\left( 2 \gamma_E - 1 -\log 4 + 2\log u\omega \right) u^2 + \cdots\right]\nn\\
&=& u\left[ \phi_{(0)}  +\left(  \underbrace{\f{1}{4}\ddot{\phi_{(0)}}}_{\psi_{(2)}} \log u^2 + \underbrace{\phi_{(2)}^{even}+ \phi_{(2)}^{odd}}_{\phi_{(2)}} \right) u^2 + \cdots\right],\label{equ:phimm31}
\eea
where $\gamma_E\simeq 0.5772$ is the Euler-Mascheroni constant and $\phi_2^{odd}$ or $\phi_2^{even}$ are respectively an odd or even function of $t$ defined by
\bea
\phi_{(2)}^{odd} &\equiv& -\f{1}{4} \int_{-\infty}^0 d\omega \sin(\omega t) \omega^2 \phi_{(0)}(\omega),\\
\phi_{(2)}^{even} &\equiv& \f{1}{4}\left( 2 \gamma_E - 1 -\log 4 \right) \ddot{\phi_{(0)}} - \f{1}{4} \int\f{d\omega}{2\pi} e^{-i\omega t}  \phi_{(0)}(\omega)\omega^2 \log \omega^2.
\eea
Here, we have assumed that $\phi_{(0)}(\omega)$ is an even function of $\omega$. 

%
%
\section{Tachyonic scalar fields coupled to gravity}\label{sec:results}
In this section, we reconstruct  the bulk geometry from the gravitational collapse of a tachyonic scalar field with $m^2 = -3$. In eq. (\ref{equ:Einstein}), we implicitly assume that the VEV of the dual operator scales as $N_c^2$. 
In this case, the back-reaction of the scalar field to the bulk geometry can not be ignored and it eventually collapses to form a black hole in the bulk. Such a collapse process is dual to the thermalization process in the boundary CFT. We study both narrow ($\f{\epsilon}{a}\lesssim1$) and broad($\f{\epsilon}{a}\gtrsim1$) waves.
\subsection{Energy conservation}
Eqs. (\ref{equ:fdot}) and (\ref{equ:fp}) give the conservation of the total energy in the bulk
\footnote{Equivalently, one has\be \f{\partial}{\partial t} \mathcal{E}^0 + \f{\partial}{\partial u} \mathcal{E}^u=0. \ee $\mathcal{E}^\alpha\equiv \left( \mathcal{E}^0, \mathcal{E}^u \right)$ is referred to as the conserved energy-momentum current in the bulk.}
\be
\f{\partial}{\partial t} M(t, u) \equiv \f{\partial}{\partial t} \int_u^\infty du \mathcal{E}^0 = \left. \mathcal{E}^u\right|^{u}_{\infty},\label{equ:MAll}
\ee
where the energy density and the energy flux along the $u$ direction are defined by
\bea
 \mathcal{E}^0 \equiv \f{2}{3u^3} \left[ f\left(V^2 + P^2\right) +\f{m^2}{u^2} \phi^2 \right],~\text{and},~\mathcal{E}^u \equiv -\f{4}{3u^3} f^2 e^{-\delta} PV = -\f{4}{3u^3} f \phi' \dot{\phi}.\label{equ:E0EuAll}
\eea
They reduce to those in eq. (\ref{equ:E0Euvac}) if the back-reaction of the scalar field to the bulk geometry is ignored. After the source is switched off, $M(t, 0)$ is the black hole mass $M_{bh}$. This can be easily seen by rewriting eq. (\ref{equ:fp}) in the form
\be
\left(\f{f-1}{u^4} \right)' =  \mathcal{E}^0.\label{equ:E0f}
\ee
Integrating over $u$, one gets from the above equation
\be
\left.M(t, 0)\right|_{t\gtrsim \Delta t} = \int_0^{\infty} du  \mathcal{E}^0 = -\lim\limits_{u\to0} \f{f-1}{u^4} = M_{bh}\equiv \left( \pi T\right)^4
\ee
by Birkhoff's theorem. Here, $T$ is the thermal equilibrium temperature of the boundary CFT or, equivalently, the Hawking temperature of the black hole formed in the bulk. 

In the boundary CFT, eq. (\ref{equ:MAll}) also gives the conservation of the boundary energy in eq. (\ref{equ:energyConservation}). Now let us take $m^2 = -3$. Near the boundary the power series solutions to the equations of motion in eq. (\ref{equ:eom}) have the form
\bea
&&\delta = \f{1}{3} \phi_{(0)}^2 u^2 +\cdots,\label{equ:deltaseries}\\
&&f = 1 + a_{(2)} u^2 + ( b_{(4)} \log u + a_{(4)} ) u^4 + \cdots,\label{equ:fseries}\\
&&\phi= u \left(\phi_{(0)} + \phi_{(2)} u^{2} + \psi_{(2)} u^{2} \log u^2  + \cdots \right)\label{equ:phiseries},
\eea
where the coefficients are given by
\bea\label{equ:a2mm03}
&&a_{(2)} = \f{2}{3} \phi_{(0)}^2,~b_{(4)} = \f{2}{3}\left( \phi_{(0)}\ddot{\phi}_{(0)} + \dot{\phi}_{(0)}^2 +  2 \phi_{(0)}^4\right),~\psi_{(2)} = \f{1}{12} \left( 3 \ddot{\phi}_{(0)}+4 \phi_{(0)}^3 \right).
\eea
Here the boundary condition in eq. (\ref{equ:phisim}) has been used and all the other coefficients of the power series can be expressed in terms of $\phi_{(0)}$ and $\phi_{(2)}$. Among them the nonsingular terms of eq. (\ref{equ:MAll}) in the limit $u\to0$ give
\be
\dot{a}_{(4)} = \frac{2}{9} \left[ 8\phi_{(0)}^3\dot{\phi}_{(0)}+6\phi_{(0)} \dot{\phi}_{(2)}+3\dot{\phi}_{(0)} \left(6 \phi_{(2)}+\ddot{\phi}_{(0)}\right) \right],
\ee
which is the time-evolution equation of  the boundary energy density
\bea\varepsilon(t) &=&-\int_{\infty}^t dt \left<\hat{O}_3 \right> \dot{\phi}_0 = \phi_{(0)} \phi_{(2)} - \f{3}{4} a_{(4)} +\f{1}{4} \dot{\phi}_{(0)}^2+ \f{1}{3}\phi_{(0)}^4.\label{equ:rhobroad}\eea
\subsection{Narrow waves$\left(\Delta t = \f{1}{\sqrt{a}} \lesssim \f{1}{T}\right)$: $\f{\epsilon}{a}\lesssim1$}
\begin{figure}
\begin{center}
\includegraphics[width=13cm]{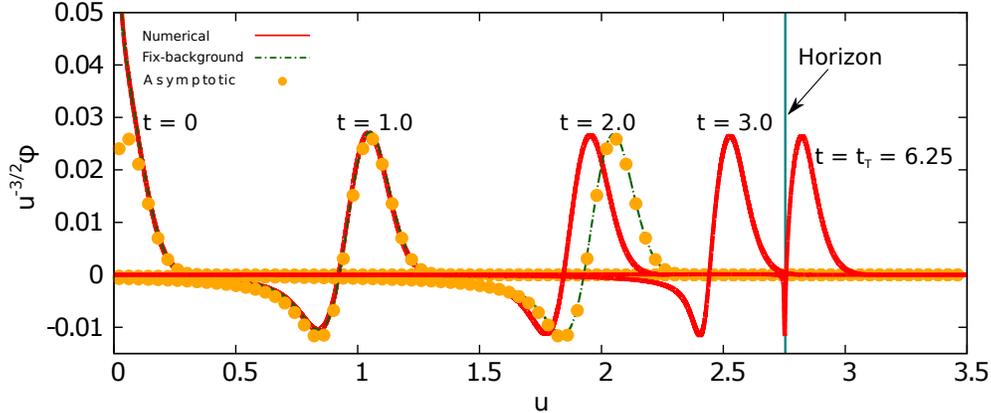}
\end{center}
\caption{A typical tachyonic narrow wave ($\epsilon = 0.5$ and $a = 60$): $\Delta t  = 0.13 \simeq \f{0.018}{T}$. This figure shows $u^{-3/2} \phi$ as a function of $u$ at different times. At $t\lesssim 1.0 = 7.7 \Delta t$, the fixed-background solution in eq. (\ref{equ:phivac}) agrees every well with our numerical solution. Then, the fixed-background solution, which can be very well approximated by the asymptotic solution in eq. (\ref{equ:phiasy}), fails to describe the collapse process later on. At $t\gtrsim t_T = 6.25$, the scalar field is (almost) completely hidden behind the apparent horizon.
}\label{fig:narrowWave}
\end{figure} 

From Fig. \ref{fig:narrowWave} one can understand how a typical narrow wave propagates in the bulk. This figure shows $u^{-\f{3}{2}}\phi$ as a function of $u$ at different times.  At $t\lesssim 7.7 \Delta t$, the fixed-background solution in eq. (\ref{equ:phivac}) is a good approximation of our numerical solution. Then, due to the back-reaction of the scalar to the bulk geometry, the group velocity of the wave starts to slow down while the fixed-background solution still propagates at the speed of light, which can be clearly seen from the asymptotic solution in eq. (\ref{equ:phiasy}). As a result, one can use the fixed-background solution to calculate the thermal equilibrium temperature $T$ and to describe the details about how the energy is injected into the CFT vacuum during $\Delta t \gtrsim t\gtrsim -\Delta t$.
\subsubsection{The thermal-equilibrium temperature $T$}
\begin{figure}
\begin{center}
\includegraphics[width=7cm]{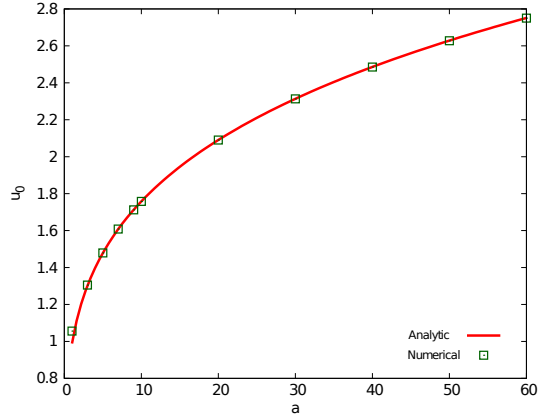}
\end{center}
\caption{The location of the apparent horizon
$u_0\equiv\f{1}{\pi T}$ for $m^2 = -3$. Here, we fix $\epsilon = 0.5$ and take $u_0$ as a function of $a$. The red solid curve (Analytic) shows the analytical result in eq. (\ref{equ:a4mm3}), which agrees very well with our numerical results for all $a\gtrsim 1$, or, equivalent, $\Delta t \lesssim \f{0.3}{T}$.
}\label{fig:u0cmp}
\end{figure} 

Let us first calculate $M_{bh} = \left(\pi T\right)^4$, the black hole mass formed in the bulk. Taking  $\hat{\epsilon}\equiv\f{\epsilon}{a}$ as a small parameter, we write $f$, $\delta$ and $\phi$ in the form\cite{Minwalla}
\bea
&&f = 1 +  \hat{\epsilon} f_1 +  \hat{\epsilon}^2 f_2 + O( \hat{\epsilon}^3),\\
&&\delta =  \hat{\epsilon} \delta_1 +  \hat{\epsilon}^2 \delta_{(2)} + O( \hat{\epsilon}^3),\\
&&\phi =  \hat{\epsilon} \phi_1 +  \hat{\epsilon}^2 \phi_2 + O( \hat{\epsilon}^3).
\eea
The equations of motion in eq. (\ref{equ:eom}) can be solved order by order in $\hat{\epsilon}$. At the first order in $\hat{\epsilon}$, 
\be
 f_1 = 0,\delta_1 =0,
\ee
and $\hat{\epsilon}\phi_1$ is given by the  fixed-background  solution in eq. (\ref{equ:phivac}). Then, the second order equation
\bea
&&\dot{f}_2 = \f{4u}{3}\phi_1' \dot{\phi}_1\label{equ:f2dot}
\eea
gives the black hole mass\footnote{Here, we take $t = \infty$ but because of the energy conservation in eq. ({\ref{equ:MAll}}) it is the black hole mass after the scalar source is switched off, that is, at time $t\gtrsim \Delta t$.}
\be
M_{bh}\equiv -a_{(4)}(\infty) = -\lim\limits_{u\to 0}\f{\hat{\epsilon}^2 f_2(\infty)}{u^4} =   -\lim\limits_{u\to 0}\f{4 \hat{\epsilon}^2}{3 u^3} \int_{-\infty}^{\infty} dt  \phi_1' \dot{\phi}_1,\label{equ:a4narrow}
\ee
where $a_{(4)}(\infty) \equiv - \f{1}{u_0^4} $ and $u_0 = \f{1}{\pi T}$ is the location of the apparent horizon. Therefore, $\left.M(t, 0)\right|_{t \gtrsim \Delta t}$ of the fixed-background solution in  eq. (\ref{equ:energyCon}) is approximately the black hole mass in the bulk. Inserting (\ref{equ:phimm31}) into (\ref{equ:a4narrow}) and only keeping the terms even in $t$, we get
\be
M_{bh}  = - \int_{-\infty}^{\infty} dt \left( 4 \dot\phi_{(0)}\phi_{(2)}^{odd}+\f{4}{3} \phi_{(0)}\dot\phi_{(2)}^{odd} \right)= -\f{8}{3} \int_{-\infty}^{\infty} dt \dot\phi_{(0)}\phi_{(2)}^{odd} = \f{4\pi \epsilon^2}{3a}\propto \f{\epsilon^2}{a}.\label{equ:a4mm3}
\ee
By comparison with our numerical results in Fig. \ref{fig:u0cmp}, we find that eq (\ref{equ:a4mm3}) is a valid approximation for $\Delta t \lesssim \f{0.3}{T}$. Using $\phi$ in eq. (\ref{equ:phim01}) as $\hat{\epsilon}\phi_1$ for the massless scalar, we have\cite{Minwalla}
\be
M_{bh} = -\f{16\pi}{3} \int_{-\infty}^{\infty} dt \dot{\phi}_{(0)} \phi_{(4)}^{odd} = \f{4\pi \epsilon^2}{3}\propto \epsilon^2,
\ee which also agrees with our numerical results in Ref. \cite{PaperI}.

The above scaling behavior of $M_{bh}$ is a consequence of the scaling symmetry. Under the scaling transformation in eq. (\ref{equ:scalingTransform}),
\be
M_{bh}\to \lambda^4 M_{bh}. 
\ee
Therefore, at $O(\hat{\epsilon}^2)$ one should has, for arbitrary $n$,
\be
M_{bh} \propto \epsilon^2 a^{n-2}=\hat{\epsilon}^2 a^n = \left(\f{\hat{\epsilon}}{\Delta t^n}\right)^2
\ee
according to the scaling transformation in eq. (\ref{equ:parametersScaling}). This scaling behavior should be independent of the shape of the scalar source.
\subsubsection{Energy injection}
Using the (approximate) analytic solution above, we discuss the procedure to obtain the boundary energy from the numerical bulk solutions in this subsection. Even though we are mainly interested in the thermalization time $t_T$ and $\varepsilon$ (strongly) depends of the shape of the boundary source, the lesson we learn from such a simplified case may be useful for studying the more complicated systems, say, in \cite{withPaul, CY01}.

\begin{figure}
\begin{center}
\includegraphics[width=7cm]{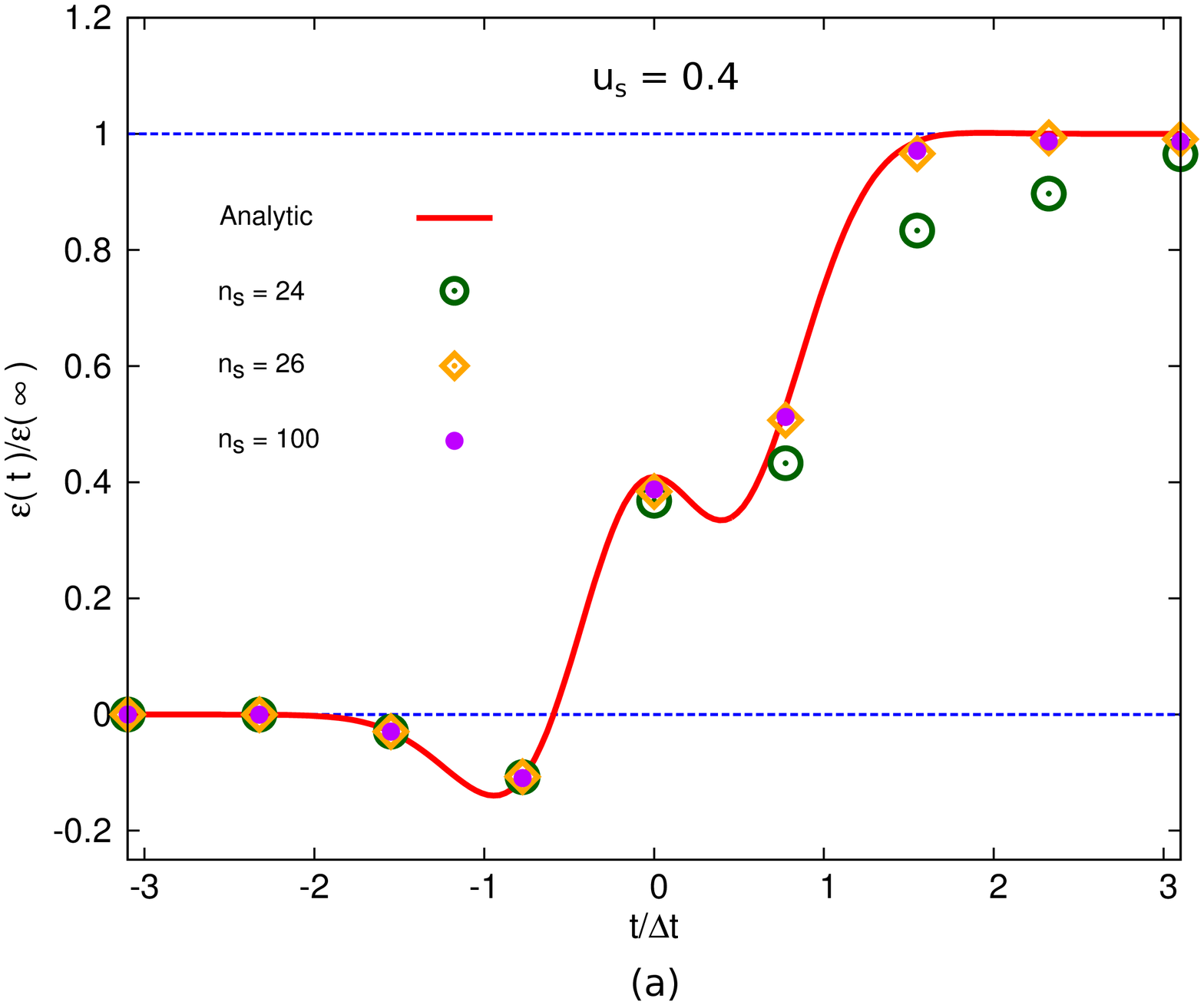}
\includegraphics[width=7cm]{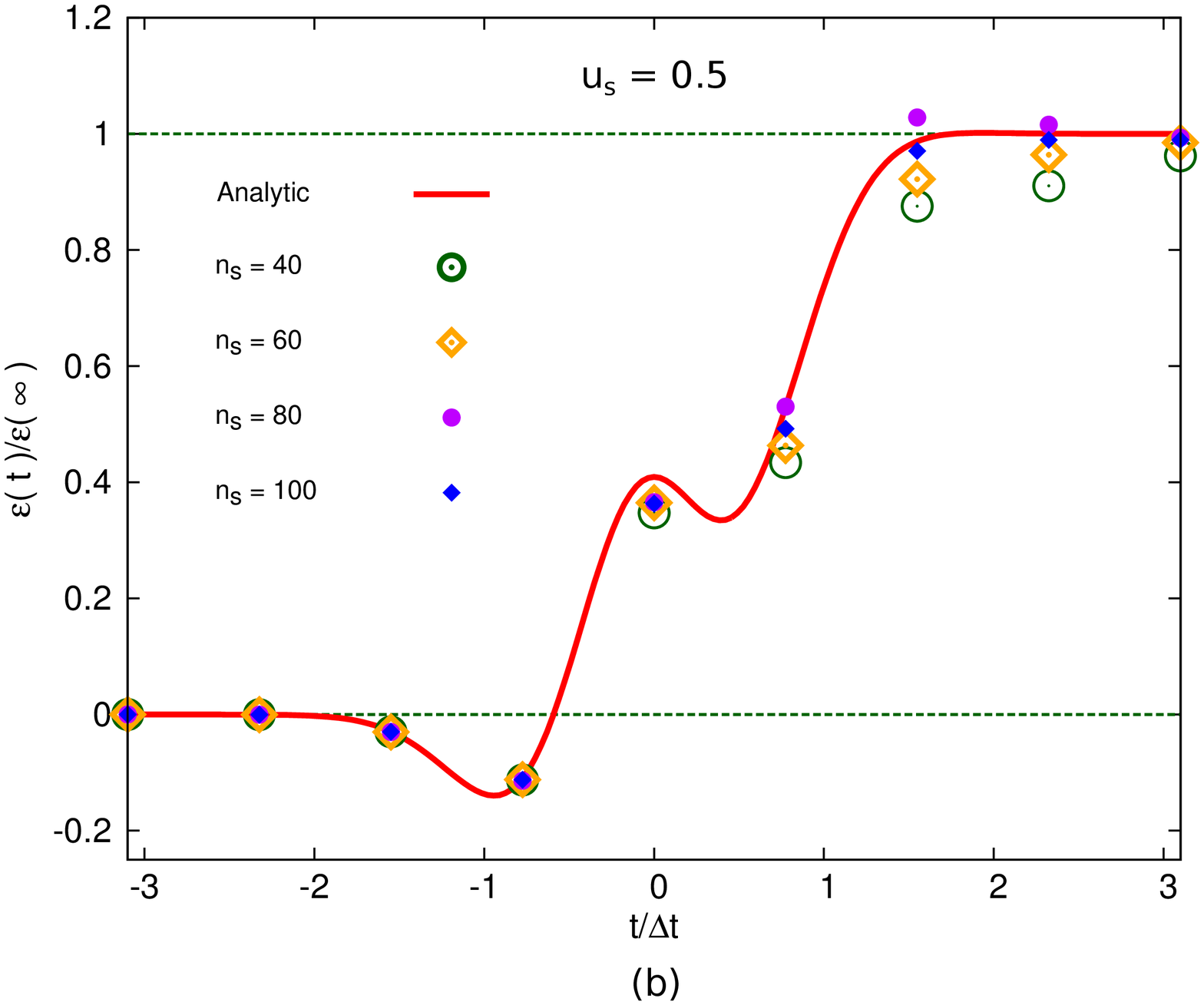}
\end{center}
\caption{The boundary energy density $\varepsilon(t)$. Fig. (a) shows the fit results of $\varepsilon$ with $u_s = 0.4$ and $n_s = 24, 26$ and $100$. We find that for $n_s\geq 26$ they do not depend on $n_s$, which agree with the result (Analytic) in eq. (\ref{equ:rho}). In contrast, for $u_s = 0.5$ the fit results always depend on $n_s$ as shown in Fig. (b). Here $(a, \epsilon) = (60, 0.5)$.
}\label{fig:t00}
\end{figure} 
The energy is injected into the vacuum in the boundary CFT according to eq. (\ref{equ:energyConservation}), that is,
\be
\varepsilon(t)=-\int_{\infty}^t dt \left<\hat{O}_3 \right> \dot{\phi}_0 = -2\int_{\infty}^t dt \dot{\phi}_{(0)} \phi_{(2)}.\label{equ:rho}
\ee
Insert into the above equation $\phi_{(2)}$ defined in eq. (\ref{equ:phimm31}), and we find the following relation for different $(a, \epsilon)$
\footnote{Note that the energy density may take negative values at some intermediate time, which can be understood from the second term on the right-hand side of  eq. (\ref{equ:t00}). One may choose a renormalization scheme different from that used in Ref. \cite{holographicReconstruction} to make $\varepsilon$ be positive all the time. The interested reader is referred to \cite{Myers:2012} for a recent discussion of holographic renormalization in the cases with $m^2 = -3$ and $-4$.}
\be
\varepsilon(t) = \f{\epsilon^2}{a}  \hat{\varepsilon}(\sqrt{a} t) - \f{1}{4} \dot{\phi}_{(0)}^2 \log a\label{equ:t00}
\ee
with
\be
\hat{\varepsilon}(t) \equiv \left.\varepsilon(t)\right|_{a=1,\epsilon=1}.
\ee
The term $\propto \log a$ is due to the conformal anomaly\cite{Skenderis:2002}.
In order to get the boundary energy density one only needs to calculate  $\phi_{(2)}$ from known solutions of eq. (\ref{equ:eom}) and then integrate over $t$ in eq. (\ref{equ:rho}).

Numerically, we find that it is easier to obtain $\varepsilon(t)$ in the following way: By inserting eq. (\ref{equ:phimm31}) into eq. (\ref{equ:f2dot}), near the boundary one has
\be
\hat{\epsilon}^2 f_2 = a_{(2)} u^2 + ( b_{(4)} \log u + a_{(4)} ) u^4 + \cdots,
\ee
where the coefficients are given by
\bea
&&a_{(2)} = \f{2}{3} \phi_{(0)}^2,~~b_{(4)} = \f{2}{3}\left( \phi_{(0)}\ddot{\phi}_{(0)} + \dot{\phi}_{(0)}^2 \right),\label{equ:b4narrow}\\
&&a_{(4)}=\f{1}{3} \left( \dot{\phi}_{(0)}^2 + 4  \phi_{(0)}\phi_{(2)} + 8 \int_{-\infty}^t dt  \dot{\phi}_{(0)}\phi_{(2)} \right).\label{equ:a4narrowt}
\eea
Accordingly,
\bea\varepsilon(t) &=& \phi_{(0)} \phi_{(2)} - \f{3}{4} a_{(4)} +\f{1}{4} \dot{\phi}_{(0)}^2,\label{equ:rhon}\eea
which is eq. (\ref{equ:rhobroad}) up to $O(\hat{\epsilon}^2)$. The coefficients $a_{(4)}$ and $\phi_{(2)}$ can be obtained by the least-squared fit of the power series expansion of $f$ and $\phi$ near the boundary to our numerical results. Here, the power series expansion of $f$ and $\phi$ takes the form
\bea\label{equ:narrowseries}
&&f_s = 1 + a_{(2)} u^2 + \sum\limits_{k=2}^{\f{n_s}{2}}\left( b_{(2k)} \log u + a_{(2k)} \right) u^{2k},\\
&&\phi_s = u\left[ \phi_{(0)} + \sum\limits_{k=1}^{\f{n_s}{2}-1}\left( \psi_{(2k)} \log u^2 + \phi_{(2k)} \right) u^{2k} \right],
\eea
where $n_s \geq 4$ is an even integer. The coefficients $\phi_{(0)}$, $\psi_{(2)}, a_{(2)}$ and $b_{(4)}$ take values according to eqs. (\ref{equ:phi0t}), (\ref{equ:phimm31}) and (\ref{equ:b4narrow}) but others are to be obtained from the least-squared fit to our numerical solutions at $u = (u_{min}, u_s)$. $u_s>u_{min}$ and $n_s$ are chosen such that the fit results of $a_{(4)}$ and $\phi_{(2)}$ do not change  by increasing $n_s$. Then, the fit result of $\varepsilon$ is obtained by eq. (\ref{equ:rhon}) or eq. (\ref{equ:rhobroad}).

Let us take for example the narrow wave with $(a, \epsilon) = (60, 0.5)$ (see Fig. \ref{fig:narrowWave}). In this case, we find that the fit results of $\varepsilon$ do not depend on $u_s$ in the range $u_s=(0.3,0.45)$.  For $u_s = 0.3, 0.4$ and $0.45$, it respectively requires $n_s \geq 18, 26$ and $42$. Fig. \ref{fig:t00}(a) shows our fit results with $u_s = 0.4$, which agree with the analytic solution in eq. (\ref{equ:rho}) for $n_s\geq 26$. However, for $u_s\gtrsim 0.5$ the fit results always depend on $n_s$ (see Fig. \ref{fig:t00}(b)), which indicates that $u = 0.5$ should be outside of the radius of convergence of the metric functions. We also find that the fit result slightly depends on $n_s$ for smaller $u_s$. It is due to the error of order $u_{min}^2 = 10^{-6}$ introduced by the boundary condition in eq. (\ref{equ:phi0}).

\subsubsection{The thermalization time}
The details about the collapse process of the tachyonic scalar is shown in Fig. \ref{fig:narrowWave}. Like the massless scalar\cite{PaperI}, it shows that the system in the boundary CFT thermalizes in a top-down manner\cite{holographicThermal}. Since the naive perturbative calculation breaks down at $t_T\gtrsim t \gg \Delta t$, we instead resort to numerical simulations to study the thermalization time $t_T$. Here $t_T$ is defined to be the first time when
\be
\min f(t_T, u) = 0.01,\label{equ:tT}
\ee
which has an interpretation in terms of a spacelike Wilson loop  $\left< W(l \simeq\f{1}{T}) \right>$ in the boundary CFT\cite{PaperI}.

\begin{figure}
\begin{center}
\includegraphics[width=7cm]{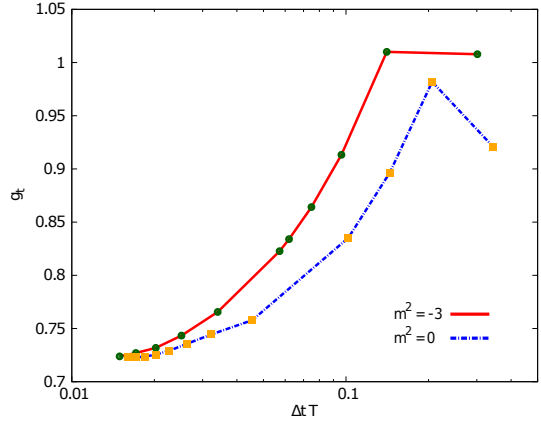}
\end{center}
\caption{The thermalization time $t_T = \f{g_t}{T}$. Here $g_t$ is taken as a function of $\Delta t~T$.  For the narrow waves with $\Delta  t \lesssim \f{0.02}{T}$, the relative difference of $g_t$ for the tachyonic and massless scalars is less than $0.8\%$. 
}\label{fig:u0}
\end{figure} 

First, let us fix $\epsilon = 0.5$ and numerically calculate $t_T$ and $T$ as functions of $a = a^{(0.5)}$, which are denoted respectively by $t_T^{(0.5)}$ and $T^{(0.5)}$. Then, let us use the scaling transformation in eq. (\ref{equ:scalingTransform}) to get the result for arbitrary $(a, \epsilon)$:
\be
\left(t_T, T, a, \epsilon \right) = \left( \f{t_T^{(0.5)}}{\lambda}, \lambda T^{(0.5)}, \lambda^2 a^{(0.5)}, 0.5 \lambda^{4-n}  \right), 
\ee
that is, the scaling parameter $\lambda$ is given by 
\be\lambda = \left( 2 \epsilon \right)^{\f{1}{4-n}},\ee
and the thermalization time $t_T$ is
\be
t_T = \f{g_t}{T}\text{~with~}g_t \equiv t_T^{( 0.5)} T^{(0.5)} = \f{t_T^{( 0.5)}}{ \pi u_0^{( 0.5)}}.
\ee
In order to compare with the results of the massless scalar in Ref. \cite{PaperI} we take $g_t$ as a function of $\Delta t ~T$. The result for $g_t$ at $0.3 \geq \Delta t~T \geq 0.015$ is shown in Fig. \ref{fig:u0}. For $\Delta  t \lesssim \f{0.02}{T}$, we find that the relative difference of $g_t$ from that of the massless scalar is less than $0.8\%$ and the mass term does not play an important role for very narrow waves. In summary, we conclude that in the cases with $\Delta t \lesssim \f{1}{T}$ the system in the boundary CFT thermalizes in a typical time $t_T\sim \f{1}{T}$.
\subsection{Broad waves($\Delta t \gtrsim \f{1}{T}$):  $\f{\epsilon}{a}\gtrsim1$}
\begin{figure}
\begin{center}
\includegraphics[width=3cm]{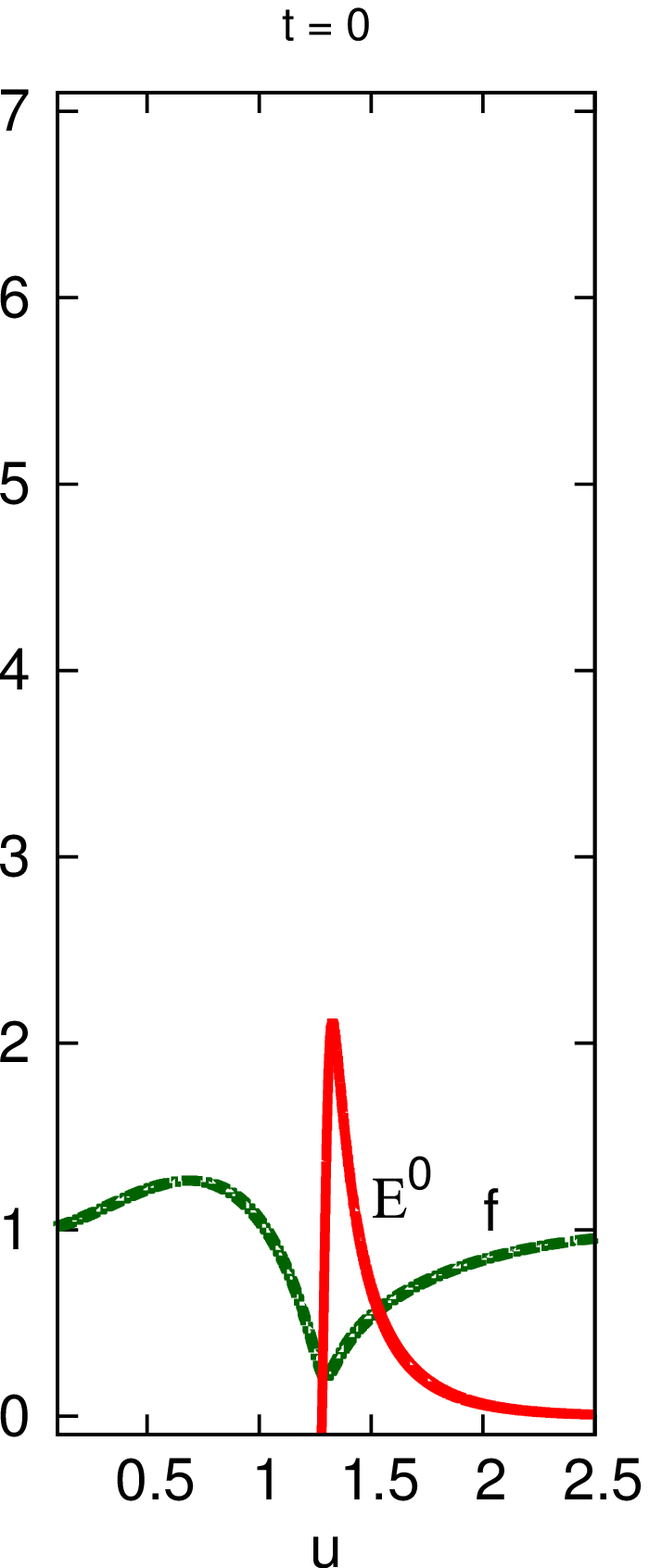}
\includegraphics[width=3cm]{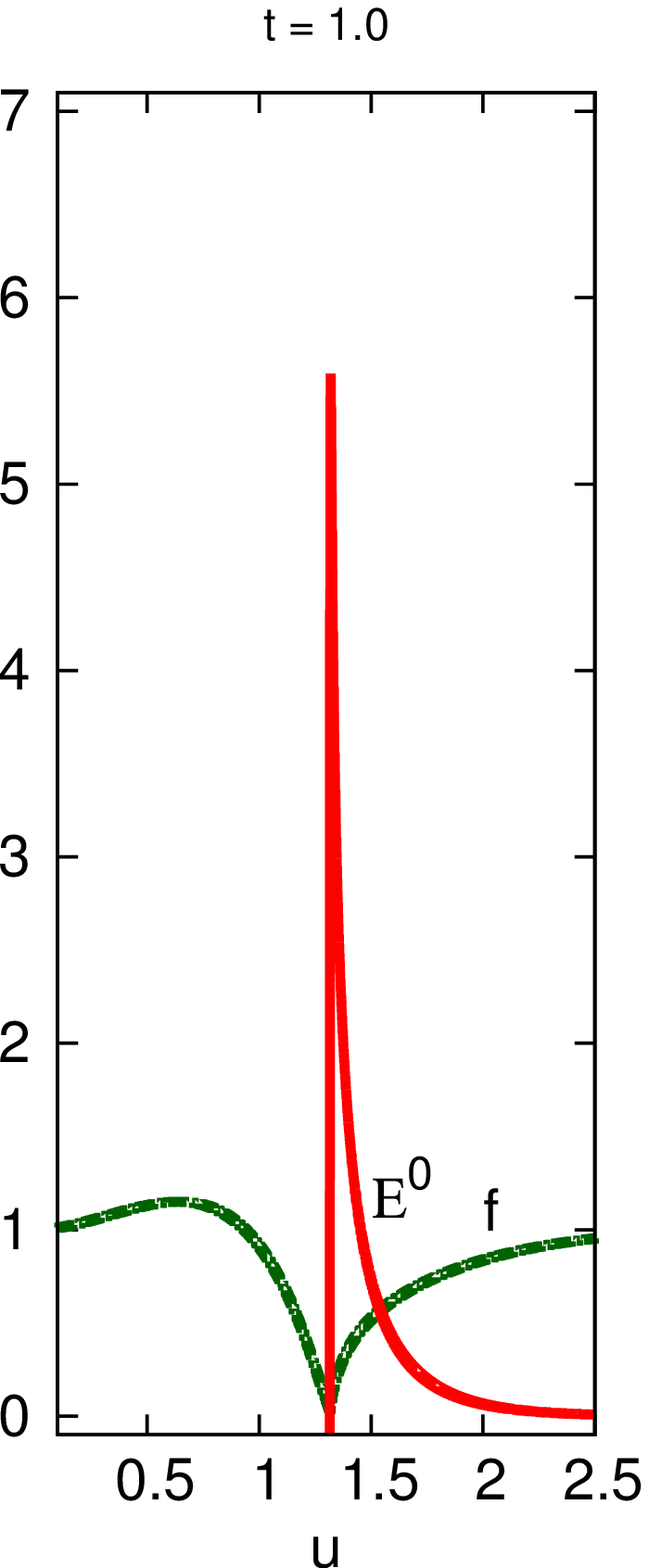}
\includegraphics[width=3cm]{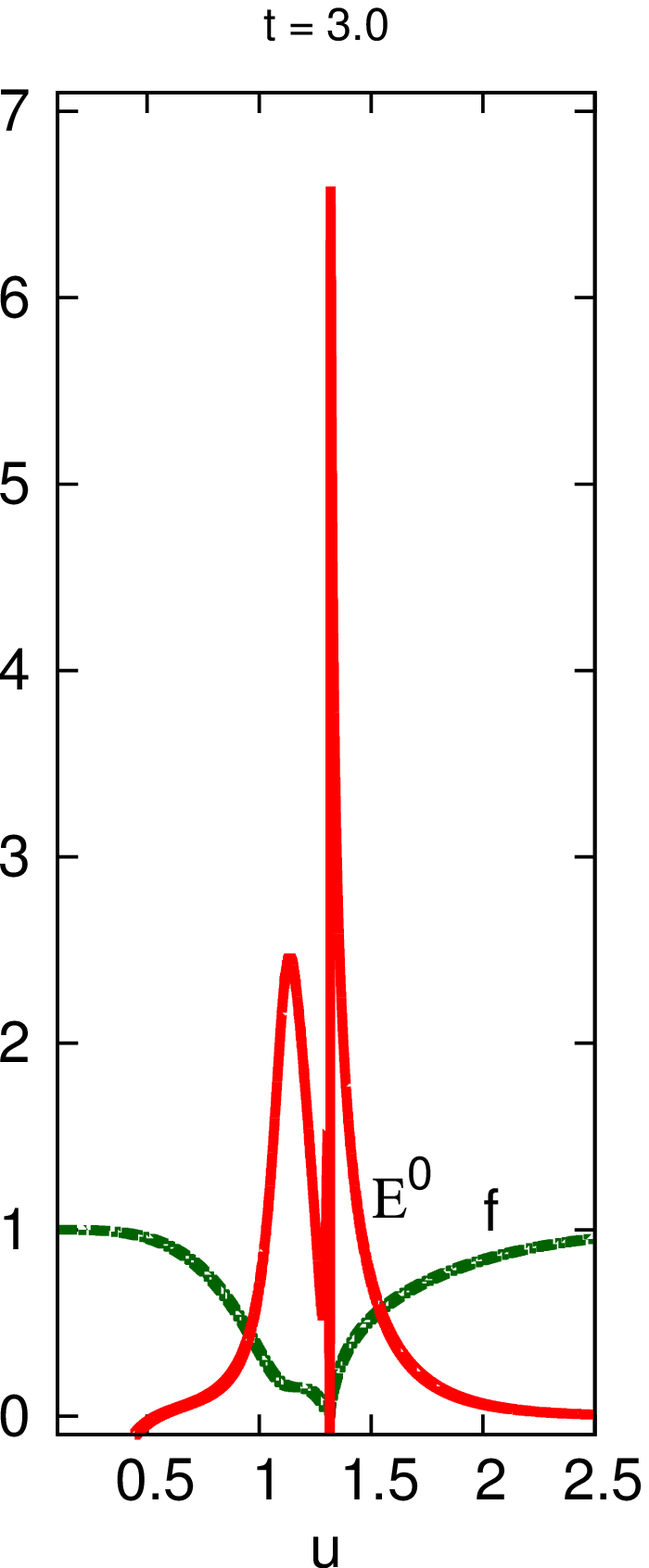}
\includegraphics[width=3cm]{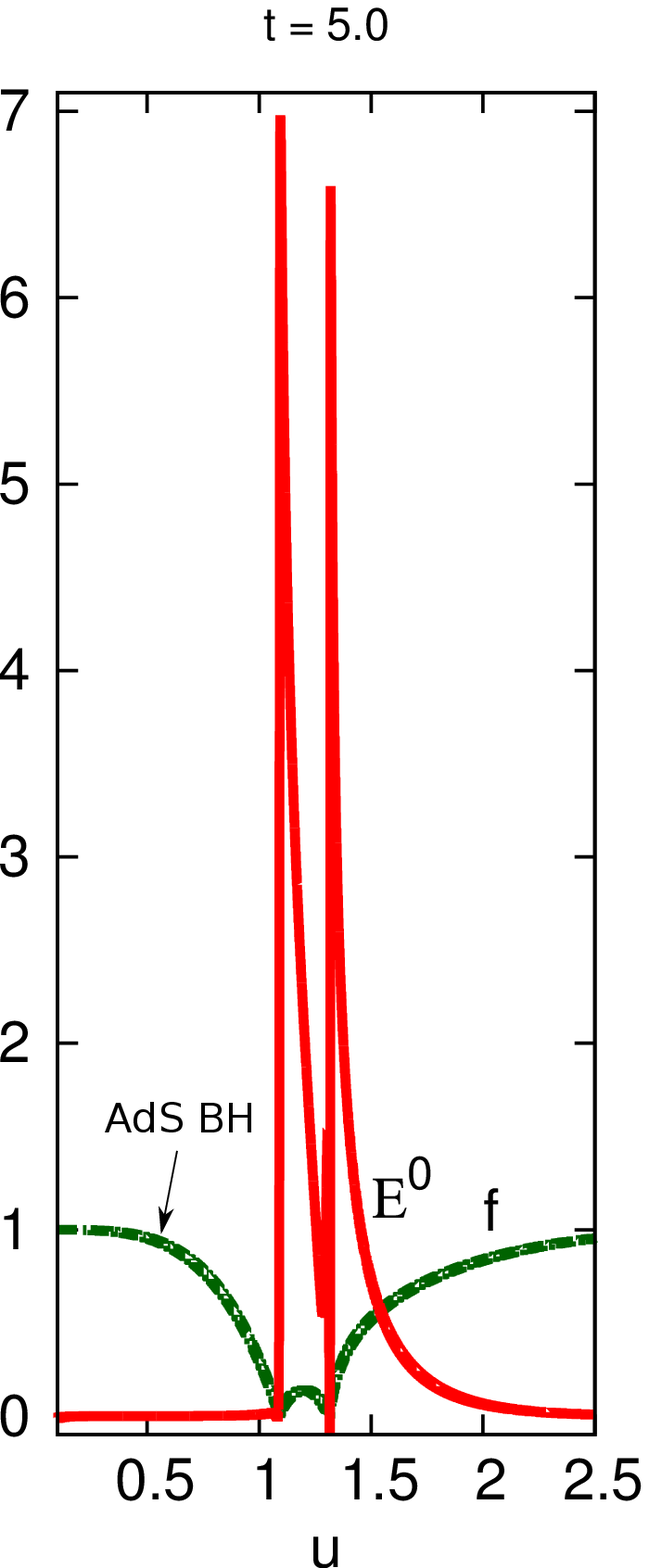}
\end{center}
\caption{Two-stage collapse in the bulk. These figures show how $\mathcal{E}^0$ and $f$ of a double-collapse solution with $(a, \epsilon) = (0.3, 0.5)$ evolve over time. The qualitative features of the time evolution of those two quantities can be understood by the conservation of $\mathcal{E}^0$ in eq. (\ref{equ:MAll}) and the back-reaction of $\mathcal{E}^0$ to $f$ in eq. (\ref{equ:E0f}).
}\label{fig:energyvsf}
\end{figure} 
At the end of this section, we briefly discuss the double-collapse solutions of the tachyonic scalar field. It also undergoes two-stage collapse in the bulk, which is qualitatively the same as the massless scalar field in Ref. \cite{PaperI}: The energy injected into the bulk from the boundary before and after $t = 0$ is respectively responsible for the two peaks in the bulk energy density $\mathcal{E}^0$. As a result, there are two local minima $f \simeq 0.01$ respectively at $u = u_L$ and $u_0$. Fig. \ref{fig:energyvsf} shows how $\mathcal{E}^0$ and $f$ of a double-collapse solution with $(a, \epsilon) = (0.3, 0.5)$ evolve over time. At each time, $f$ reaches its minima around the locations of the peaks in $\mathcal{E}^0$. And the unsmoothness near its minima can be qualitatively understood by eq. (\ref{equ:E0f}): Near the peaks,
\be
\mathcal{E}^0=\left(\f{f-1}{u^4} \right)' \simeq \f{f'}{u^4}.
\ee
At $t\gtrsim \Delta t$, the total bulk energy $M(t,0) = M_{bh}$ is conserved according to eq. (\ref{equ:MAll}), which governs the time evolution of $\mathcal{E}^0$ shown in Fig. \ref{fig:energyvsf}.

\begin{figure}
\begin{center}
\includegraphics[width=7cm]{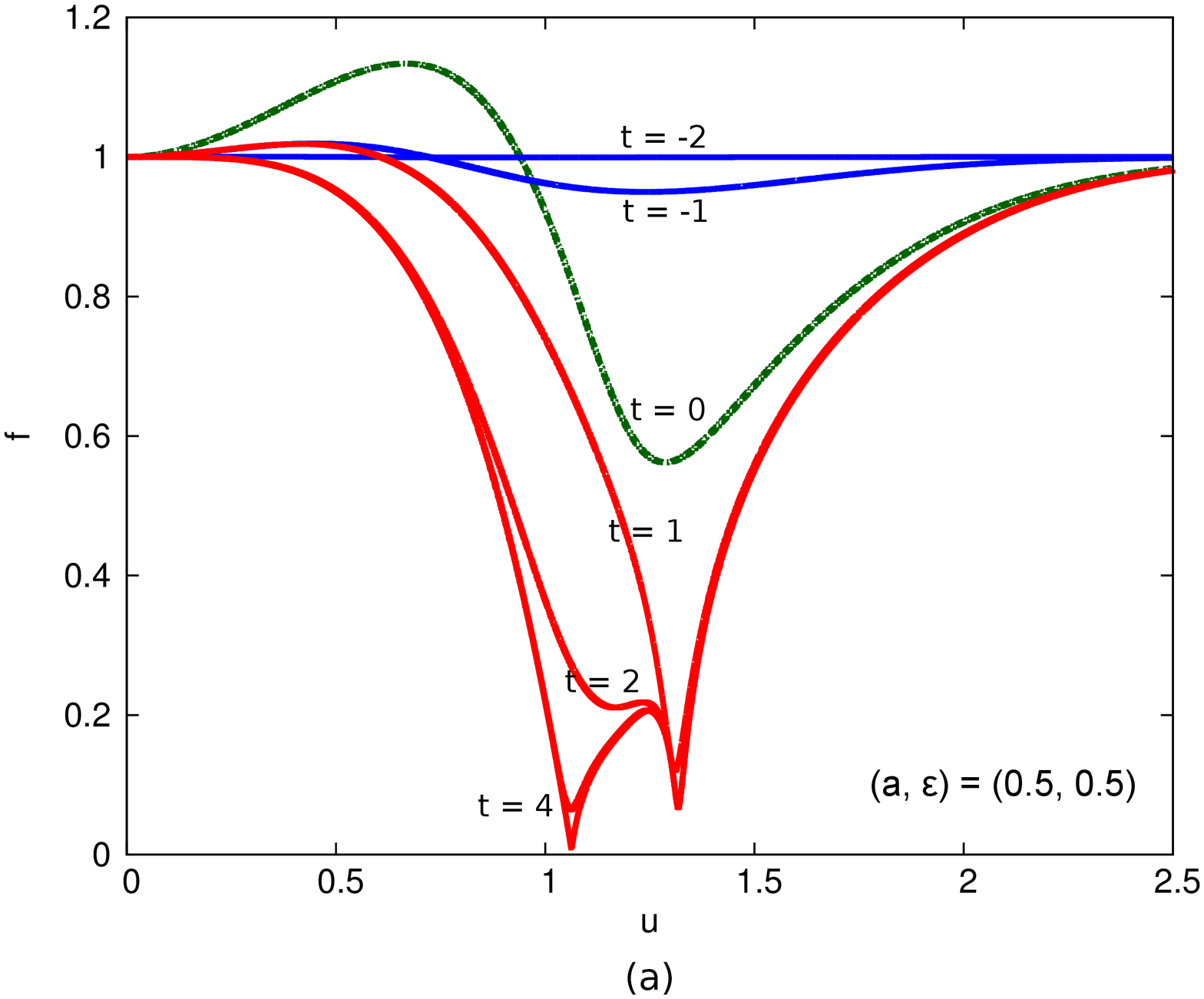}
\includegraphics[width=7cm]{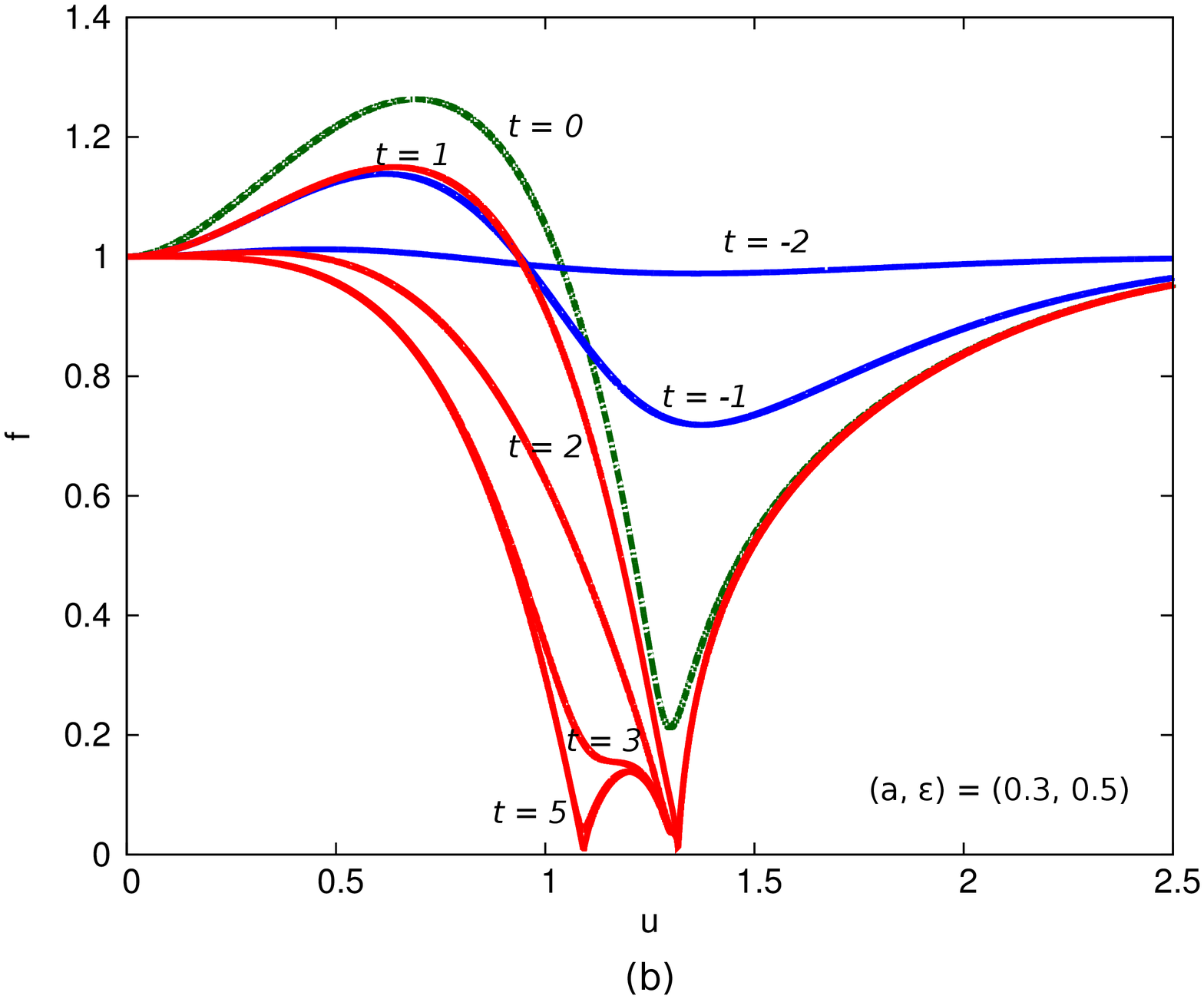}
\end{center}
\caption{Single-collapse (Fig. (a)) and double-collapse (Fig.(b)) solutions. They are respectively the  analogs of the single-collapse and double-collapse solutions of the massless scalar fields in Ref. \cite{PaperI}.
}\label{fig:broadf}
\end{figure} 

However, as shown in Figs. \ref{fig:energyvsf} and \ref{fig:broadf}, $f$  at $t = 0$  is far from that of the AdS black hole metric near the boundary\footnote{For $u_{min}=0.0001/0.001$, we find the fit result of $\varepsilon$ always depends on $n_s$ and $u_s$. It indicates that even $u_s\sim0.1$ is outside of the radius of convergence of the metric functions. The calculation with a smaller $u_{min}$ is numerically too demanding for us.}. This is in sharp contrast with the double-collapse solutions of the massless scalar field. Such a difference can be easily understood according to eqs. (\ref{equ:a2mm03}) and (\ref{equ:a2m00}), which show that $f \simeq 1 + \f{2}{3} \phi_{(0)}^2 u^2$ for $m^2 = -3$ while $f \simeq 1 - \f{1}{3} \dot\phi_{(0)}^2 u^2$ for $m^2 = 0$. At $t = 0$ the system does not look like thermalized for either local or non-local operators.

To conclude this subsection, we discuss the thermalization time for broad waves. In this case, we define the thermalization time $t_T$ by subtracting out $\Delta t$ from that defined in eq. (\ref{equ:tT}). That is, we study the thermalization of the isolated system after the source is switched off. For the solution with $(a, \epsilon) = (0.3, 0.5)$, we find $t_T =1.85  = \f{0.83}{T}$ with $u_0 = 1.09$. We find that this is also parametrically true for other double-collapse solutions. Therefore, we conclude that the typical thermalization time for such a strongly coupled system is always of order $\f{1}{T}$ no matter whether the non-equilibrium initial state is produced by marginal or relevant perturbations.
%
%
%
\section{Conclusions}\label{sec:discussion}
In this paper, we study the thermalization process of a strongly coupled system via AdS/CFT. The non-equilibrium initial state of the system is created  by turing on a scalar source during $\Delta t \gtrsim t \gtrsim -\Delta t$, which is coupled to a relevant composite operator. On the gravity side, we study the gravitational collapse of a tachyonic scalar field with $m^2 = -3$ in $AdS_5$. In the case with $\Delta t \lesssim \f{0.3}{T}$ we find that the black hole mass $M_{bh}\equiv \left(\pi T \right)^4\propto \left(\f{\hat{\epsilon}}{\Delta t}\right)^2$ with $\hat{\epsilon}\equiv \f{\epsilon}{a}$ being the amplitude of the scalar source. In this case, the injection of energy into the CFT vacuum can be very well described by perturbative calculations. Moreover, for $\Delta t < \f{0.02}{T}$, the thermalization time $t_T$ is found to be quantitatively the same as that of the non-equilibrium state created by the marginal perturbation in Ref. \cite{PaperI}. For the case with $\Delta t \gtrsim \f{1}{T}$ a non-equilibrium intermediate state at $t\simeq 0$ is found in double-collapse solutions, which is very different from that of the massless scalar field. In all the cases we find that the system, after the scalar source is turned off,  thermalizes in a typical time $t_T \simeq \f{O(1)}{T}$. Such a rapid thermalization seems typical of such a strongly coupled CFT (see \cite{JanikPeschanski2005,Yuri:2008, GrumillerPaul, Gubser:2008pc,AlvarezGaume:2008fx,CY01, Lin:2009pn, Yuri:2009,Kovchegov:2009du, CY02,Heller:2011, KiritsisTaliotis,Gubser:2012,Heller:2012km,Taliotis:2012sx} for various other non-equilibrium initial states).

After the source is switched off, one has
\be
\mathcal{E}^u(t,0) = 0 = \mathcal{E}^u(t,\infty).
\ee
Both the boundary at $u=0$ and the hypersurface at $u=\infty$ act as reflecting mirrors to the scalar field. As a result, its total energy $M(t,0)$ is conserved. In our case, as shown in Fig. \ref{fig:narrowWave}, the scalar field always propagates inwards and eventually hides behind the apparent horizon. It can not reach $u=\infty$. In contrast, in the global coordinate the scalar field can travel from the boundary/center to the center/boundary in finite time. In this case, both the center and the boundary act as reflecting mirrors. The scalar field may collapse to form a black hole after oscillating between the boundary and the center several times\cite{Bizon01}. The conserved energy-momentum current $\mathcal{E}^\alpha\equiv(\mathcal{E}^0, \mathcal{E}^x)$ in global $AdS_4$ is defined in eq. (\ref{equ:E0EuGlobal}), which is useful for understanding the qualitative features of such a collapse process (see Fig. \ref{fig:global}). A more comprehensive description in terms of $\mathcal{E}^\alpha$ would help understand better such collapse processes in global $AdS$. Besides, many discussions in Secs. \ref{sec:eom} and \ref{vacuum} are also applicable to massive scalar fields (with $m^2 >0$), which are relevant for the thermalization of the non-equilibrium state created by irrelevant perturbations. However, $\phi$ is divergent near the boundary $u = 0$ (see eq. (\ref{equ:phisim})) and the back-reaction of the matter field to the bulk geometry requires a more careful analysis\cite{Witten,holographicReconstruction}. Moreover, in our double-collapse solutions, at $t=0$ the speed of light-like geodesics $\f{du}{dt}\equiv f e^{-\delta}\simeq 1 + \f{1}{3}\phi_{(0)}^2 u^2 > 1$ at $u\sim 0$. It would be interesting to learn more details about such an intermediate state as well as the thermalization process by studying other probes such as heavy quarks\cite{energylossandpT}, dileptons\cite{Baier:2012a} or prompt photons\cite{Baier:2012b}. We leave those unanswered questions for future research.

\section*{Acknowledgements}
The author would like to thank P. Romatschke, S. Stricker and A. Vuorinen for reading this manuscript and providing illuminating comments. The author would also like to thank F. Gelis and G. Soyez for inspiring discussions of some numerical methods. This work is supported by the Agence Nationale de la Recherche project \#~11-BS04-015-01.

\appendix
%
%
\section{Near-boundary behavior of the massless scalar field}\label{sec:appA}
For the massless scalar field, $\Delta = 4$ and $n = 2$. The power series expansion of the fixed-background solution in eq. (\ref{equ:phivac}) near the boundary is as follows
\bea\phi &=&  \RE \int\f{d\omega}{2\pi} e^{-i\omega t} \phi_{(0)}(\omega)\left[1 + \f{\omega^2 u^2}{4} + \f{\omega^4 u^4}{64}\left( 3 - 4 \gamma_E  - 4\log\f{\omega u}{2} \right)+ \cdots \right] \nn\\
&=& \phi_{(0)}  - \f{u^2}{4}\ddot{\phi_{(0)}}  + \left( \underbrace{\phi_{(4)}^{even}+ \phi_{(4)}^{odd}}_{\phi_{(4)}} + \psi_{(4)} \log u^2 \right) u^4 + \cdots,\label{equ:phim01}\eea
where 
\bea
\psi_{(4)}&\equiv& -\f{\phi_{(0)}^{(4)}}{32},\phi_{(4)}^{odd} \equiv -\int_{-\infty}^0 \f{d\omega}{32\pi} \sin(\omega t) \omega^4 \phi_{(0)}(\omega),\\
\phi_{(4)}^{even} &\equiv& \f{\phi_{(0)}^{(4)}}{64}\left( 3 - 4 \gamma_E + 4 \log 2 \right) - \f{1}{32} \int\f{d\omega}{2\pi} e^{-i\omega t}  \phi_{(0)}(\omega)\omega^4 \log \omega^2.\eea

Near the boundary, the power series solutions to the equations of motion in eq. (\ref{equ:eom}) take the form\footnote{At $t = 0$, $f = 1 + \left( \f{2 \epsilon^2}{3} \log u + a_{(4)} \right) u^4 + \cdots$. For the double-collapse solutions, one always has $|a_{4}| \gg \left|\f{2\epsilon^2}{3} \log u\right|$ at $u_L \geq u \geq u_{min}$. This makes $f$ look like that of the AdS black hole at $t = 0$.}
\bea
&&f = 1 + a_{(2)} u^2 + ( b_{(4)} \log u + a_{(4)} ) u^4 + \cdots,\\
&&\phi= \phi_{(0)} + \phi_{(2)} u^{2} +  \phi_{(4)} u^{4} + \psi_{(4)} u^{4} \log u^2  + \cdots,\\
&&\delta =\frac{1}{3} u^2 \dot{\phi}_{(0)}^2+\frac{1}{72} u^4 \left(16 \dot{\phi}_{(0)}^4+3 \ddot{\phi}_{(0)}^2-6 \dot{\phi}_{(0)} \phi_{(0)}^{(3)} \right),
\eea
where the coefficients are given by
\bea\label{equ:a2m00}
&&a_{(2)} = -\f{1}{3} \dot{\phi}_{(0)}^2,~b_{(4)} = \f{1}{6}\left(   4 \dot{\phi}_{(0)}^4 +  \ddot{\phi}_{(0)}^2 - 2 \dot{\phi}_{(0)} \phi^{(3)}_{(0)} \right),\\
&&\phi_{(2)} = -\f{1}{4} \ddot{\phi}_{(0)},~\psi_{(4)} = -\f{1}{32} \left(  \phi_{(0)}^{(4)} - 8 \dot{\phi}_{(0)}^2 \ddot{\phi}_{(0)} \right).
\eea
The nonsingular terms of eq. (\ref{equ:MAll}) in the limit $u\to0$ gives
\be
\dot{a}_4 = \frac{1}{36} \left(192 \phi_{(4)}\dot{\phi}_{(0)}+32\dot{\phi}_{(0)}^3\ddot{\phi}_{(0)}+6\ddot{\phi}_{(0)}\phi_{(0)}^{(3)}-3\dot{\phi}_{(0)}\phi_{(0)}^{(4)}\right).
\ee
As a result, we have
\bea\varepsilon(t) &=&-\int_{\infty}^t dt \left<\hat{O}_4 \right> \dot{\phi}_0 = -4 \int_{-\infty}^t dt \dot{\phi}_{(0)} \phi_{(4)}\nn\\
& =& - \f{3}{4} a_{(4)} +\f{3}{32} \ddot{\phi}_{(0)}^2 - \f{1}{16} \dot{\phi}_{(0)} \phi_{(0)}^{(3)}+ \f{1}{6} \dot{\phi}_{(0)}^4.\label{equ:rhobroadm0}\eea
%
%
\section{Gravitational collapse of massless scalars in global $AdS_4$}\label{sec:appB}
%
%
%
\begin{figure}
\begin{center}
\includegraphics[width=7cm]{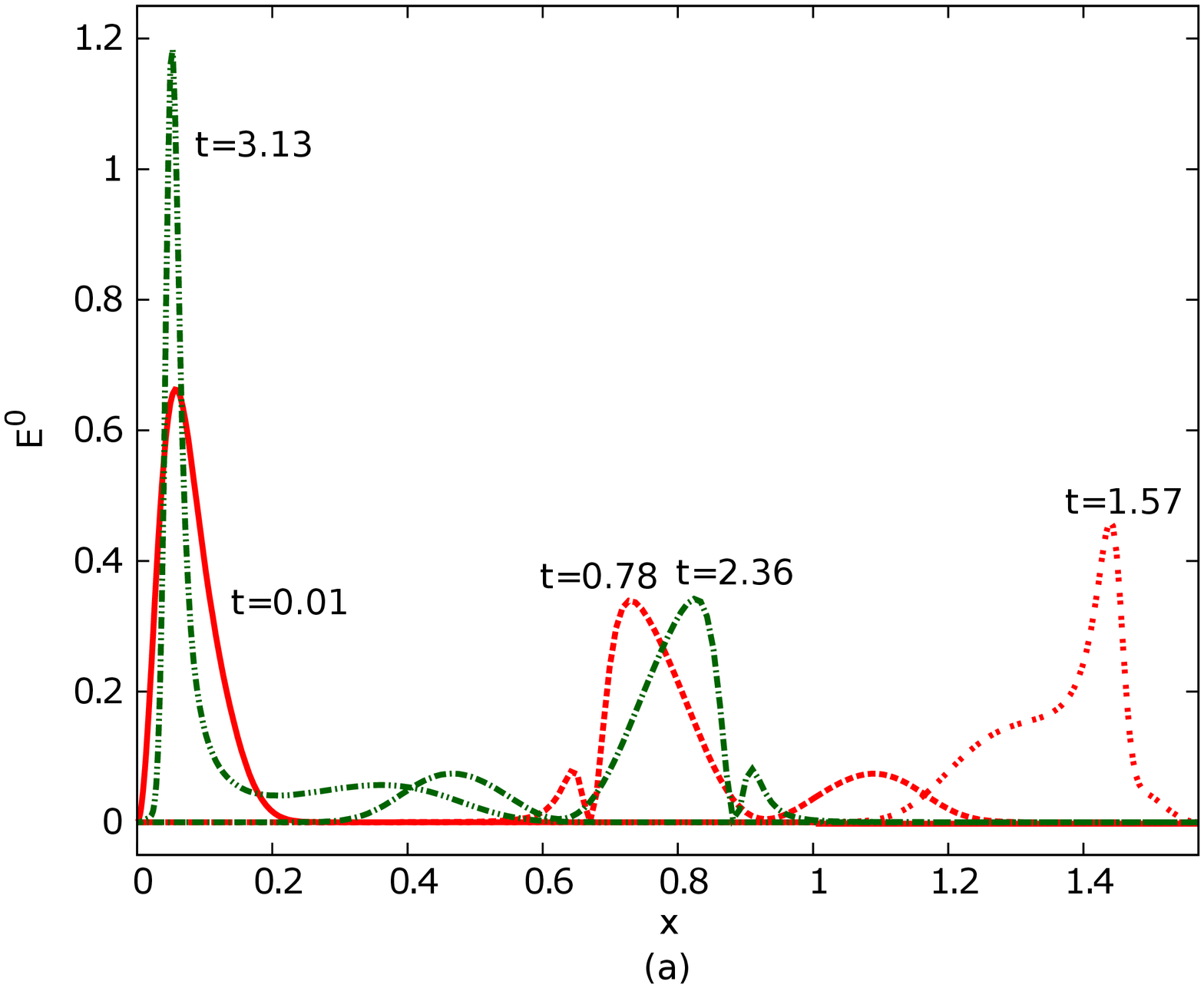}
\includegraphics[width=7cm]{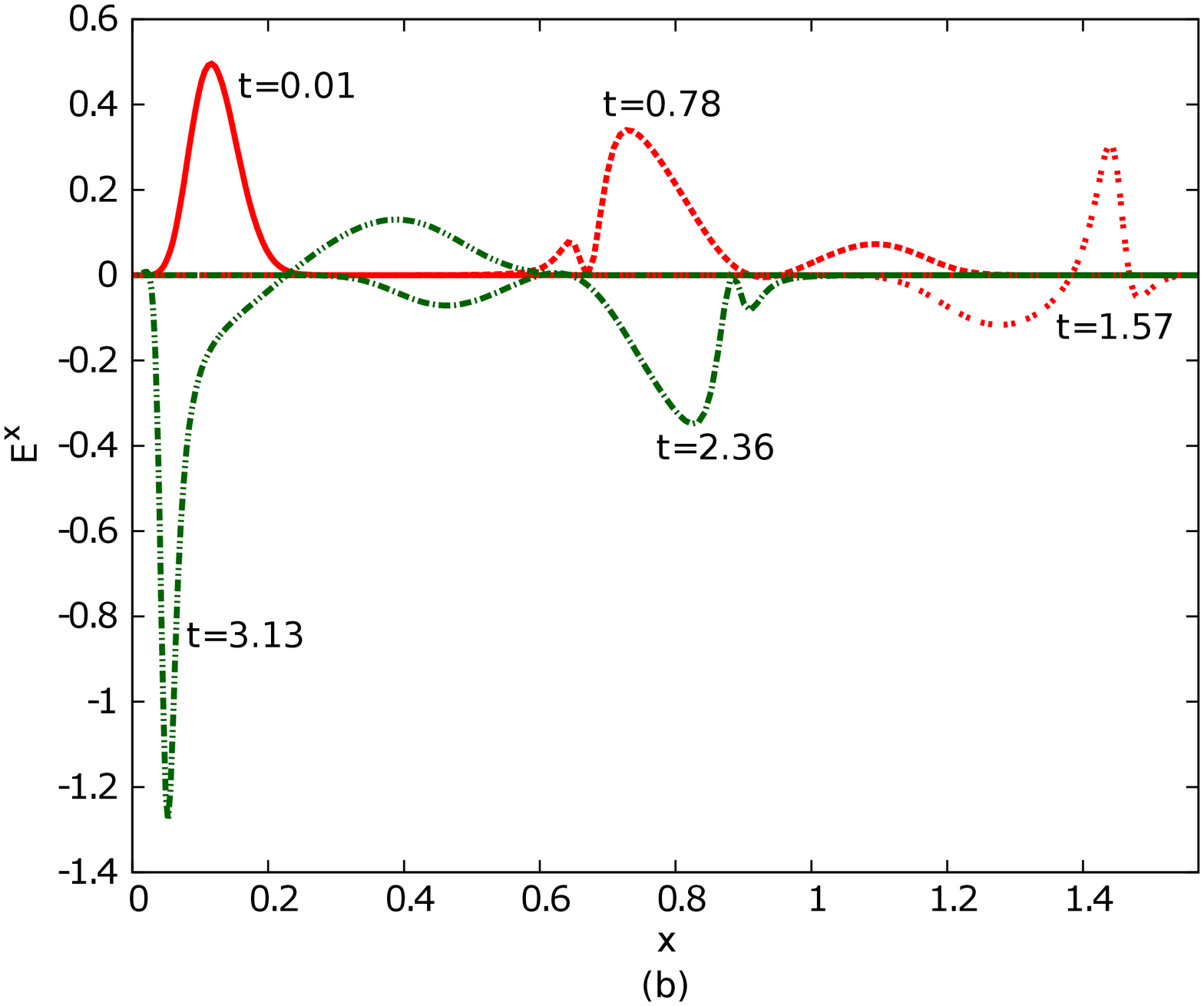}
\end{center}
\caption{Gravitational collapse of a massless scalar in global $AdS_4$. Here, we show the time evolution of $\mathcal{E}^0$ (Fig. (a)) and $\mathcal{E}^x$ (Fig. (b)) of a solution which collapses to form a black hole at $t \sim \pi$. At $t = 0$, the matter field locates mostly near the center ($x = 0$). Then, the matter field propagates upwards as a result from the reflection by the center. Afterwards, it keeps going up before reaching the boundary, which forces it to fall inwards. At $t\sim\pi$, $\mathcal{E}^0$ is more sharply peaked near $x = 0$ than that at $t = 0$. The force from the reflection can not resist the gravity and the scalar field will eventually collapse to form a black hole.   
}\label{fig:global}
\end{figure} 

The metric of $AdS_4$ in the global coordinate takes the form\cite{Bizon01}
\be
ds^2 = \f{1}{\cos^2x}\left( -f e^{-2\delta} dt^2 + f^{-1}dx^2 + \sin^2x d\Omega^2 \right),
\ee
where the metric functions $f$ and $\delta$ are functions of $t$ and $x$. Here, the boundary of $AdS_4$ locates at $x = \f{\pi}{2}$. It takes only $t = \f{\pi}{2}$ for light to travel from the boundary/center to the center/boundary. From eqs. (\ref{equ:KleinGolden}) and (\ref{equ:Einstein}), one has
\begin{subequations}\label{equ:eomglobal}
\bea
&&\f{\partial}{\partial t}M(t,x_1,x_2)\equiv\f{\partial}{\partial t}\int_{x_1}^{x_2}\mathcal{E}^0 = \mathcal{E}^x(t,x_1)-\mathcal{E}^x(t,x_2),\label{equ:MGlobal}\\
&&\dot{V} = \f{1}{\tan^2x} \left(\tan^2x f e^{-\delta} P \right)^\prime,\dot{P} = \left( f e^{-\delta} V \right)^\prime,\label{equ:phiGlobal}\\
&&\delta^\prime=-\sin x\cos x \left(  V^2 + P^2 \right),\dot{f} = \f{\cos^3 x}{\sin x}\mathcal{E}^x,\label{equ:metricGlobal}
\eea
\end{subequations}
where the derivatives with respect to $t$ and $x$ are denoted respectively by overdots and primes, $P\equiv \phi^\prime$, $V\equiv f^{-1} e^\delta \dot{\phi}$ and the conserved energy-momentum current $\mathcal{E}^\alpha \equiv \left(\mathcal{E}^0, \mathcal{E}^x\right)$ is defined by
\bea
 \mathcal{E}^0 \equiv \tan^2x f\left(V^2 + P^2\right),~\text{and},~\mathcal{E}^x \equiv -2 \tan^2x f^2 e^{-\delta} PV.\label{equ:E0EuGlobal}
\eea
After the boundary source is switched off, one has
\be
P(t,0)=0,\text{~and~}V\left(t,\f{\pi}{2}\right) = 0.
\ee
As a result, the total energy $M(t,0,\f{\pi}{2})$ is a constant according to eq. (\ref{equ:MGlobal}). $\mathcal{E}^0$ and $\mathcal{E}^x$ help understand some details on how the scalar field evolves over time in the bulk. Fig. \ref{fig:global} shows the time evolution of $\mathcal{E}^0$ and $\mathcal{E}^x$ of a solution with the initial condition given by $P(0,x) = 0$ and $V(0,x) = \f{80}{\pi}\exp\left(-\f{ 1024 \tan^2x}{\pi^2} \right)$. It illustrates the general features about such a collapse process. All the discussions here can be easily generalized to the cases of massive/tachyonic scalar fields in global $AdS_{d+1}$ or in Minkowski space (by taking the limit $x\to 0$).

\end{document}